

Instructions:  Postscript file appended after "bye" command


\font\titlefont = cmr10 scaled\magstep 4
\font\sectionfont = cmr10
\font\littlefont = cmr5
\font\eightrm = cmr8

\def\sss{\scriptscriptstyle}

\magnification = 1200

\global\baselineskip = 1.2\baselineskip
\global\parskip = 4pt plus 0.3pt
\global\abovedisplayskip = 18pt plus3pt minus9pt
\global\belowdisplayskip = 18pt plus3pt minus9pt
\global\abovedisplayshortskip = 6pt plus3pt
\global\belowdisplayshortskip = 6pt plus3pt


\def\endignore{}
\def\ignore #1\endignore{}

\newcount\dflag
\dflag = 0


\def\monthname{\ifcase\month
\or Jan \or Feb \or Mar \or Apr \or May \or June%
\or July \or Aug \or Sept \or Oct \or Nov \or Dec
\fi}

\def\timestring{{\count0 = \time%
\divide\count0 by 60%
\count2 = \count0
\count4 = \time%
\multiply\count0 by 60%
\advance\count4 by -\count0
\ifnum\count4 < 10 \toks1 = {0}
\else \toks1 = {} \fi%
\ifnum\count2 < 12 \toks0 = {a.m.}
\else \toks0 = {p.m.}
\advance\count2 by -12%
\fi%
\ifnum\count2 = 0 \count2 = 12 \fi
\number\count2 : \the\toks1 \number\count4%
\thinspace \the\toks0}}

\def\timestamp{\number\day\ \monthname\ \number\year\quad\timestring}

\def\today{\ifcase\month\or January\or February\or March\or
 April\or May\or June\or July\or August\or September\or
 October\or November\or December\fi \space\number\day, \number\year}


\def\draftmode{\global\dflag = 1
\headline{Preliminary Draft \hfil \timestamp}}


\def\endtitle{}
\def\title#1\endtitle{\vskip.5in\titlefont
\global\baselineskip = 2\baselineskip
#1\vskip.4in
\baselineskip = 0.5\baselineskip\rm}

\def\endauthors{}
\def\authors#1\endauthors{#1}

\def\endabstract{}
\def\abstract#1\endabstract{\vskip .3in%
\centerline{\sectionfont\bf Abstract}%
\vskip .1in
\noindent#1}

\newcount\nsection
\newcount\nsubsection

\def\section#1{\global\advance\nsection by 1
\nsubsection=0
\bigskip\noindent\centerline{\sectionfont \bf \number\nsection.\ #1}
\bigskip\rm\nobreak}

\def\subsection#1{\global\advance\nsubsection by 1
\bigskip\noindent\sectionfont \sl \number\nsection.\number\nsubsection)\
#1\bigskip\rm\nobreak}

\def\topic#1{{\medskip\noindent $\bullet$ \it #1:}}
\def\endtopic{\medskip}

\def\appendix#1#2{\bigskip\noindent%
\centerline{\sectionfont \bf Appendix #1.\ #2}
\bigskip\rm\nobreak}


\newcount\nref
\global\nref = 1

\def\ref#1#2{\xdef #1{[\number\nref]}
\ifnum\nref = 1\global\xdef\therefs{\noindent[\number\nref] #2\ }
\else
\global\xdef\oldrefs{\therefs}
\global\xdef\therefs{\oldrefs\vskip.1in\noindent[\number\nref] #2\ }%
\fi%
\global\advance\nref by 1
}

\def\listrefs{\vfill\eject\section{References}\therefs}


\newcount\nfoot
\global\nfoot = 1

\def\foot#1#2{\xdef #1{(\number\nfoot)}
\footnote{${}^{\number\nfoot}$}{\eightrm #2}
\global\advance\nfoot by 1
}


\newcount\nfig
\global\nfig = 1

\def\fig#1{\xdef #1{(\number\nfig)}
\global\advance\nfig by 1
}


\newcount\cflag
\newcount\nequation
\global\nequation = 1
\def\eqlabel{(1)}

\def\nexteqno{\ifnum\cflag = 0
\global\advance\nequation by 1
\fi
\global\cflag = 0
\xdef\eqlabel{(\number\nequation)}}

\def\lasteqno{\global\advance\nequation by -1
\xdef\eqlabel{(\number\nequation)}}

\def\label#1{\xdef #1{(\number\nequation)}
\ifnum\dflag = 1
{\escapechar = -1
\xdef\draftname{\littlefont\string#1}}
\fi}

\def\clabel#1#2{\xdef\eqlabel{(\number\nequation #2)}
\global\cflag = 1
\xdef #1{\eqlabel}
\ifnum\dflag = 1
{\escapechar = -1
\xdef\draftname{\string#1}}
\fi}

\def\cclabel#1#2{\xdef\eqlabel{#2)}
\global\cflag = 1
\xdef #1{\eqlabel}
\ifnum\dflag = 1
{\escapechar = -1
\xdef\draftname{\string#1}}
\fi}


\def\eeq{}

\def\eqnn #1\eeq{$$ #1 $$}

\def\eq #1\eeq{\xdef\draftname{\ }
$$ #1
\eqno{\eqlabel \rlap{\ \draftname}} $$
\nexteqno}

\def\eqnumber{\eqlabel}


\def\eol{& \eqlabel \rlap{\ \draftname} \crcr
\nexteqno
\xdef\draftname{\ }}

\def\eeol{& \eqlabel \rlap{\ \draftname}
\nexteqno
\xdef\draftname{\ }}

\def\eolnn{\cr
\global\cflag = 0
\xdef\draftname{\ }}

\def\eeolnn{\xdef\draftname{\ }}

\def\eqa #1\eeq{\xdef\draftname{\ }
$$ \eqalignno{ #1 } $$
\global\cflag = 0}


\def\ie{{\it i.e.\/}}
\def\eg{{\it e.g.\/}}
\def\etc{{\it etc.\/}}
\def\etal{{\it et.al.\/}}
\def\apriori{{\it a priori\/}}
\def\aposteriori{{\it a posteriori\/}}
\def\via{{\it via\/}}
\def\vs{{\it vs.\/}}
\def\cf{{\it c.f.\/}}
\def\adhoc{{\it ad hoc\/}}
\def\bll{$\bullet$}

\def\myinstitution{
   \centerline{\it Physics Department, McGill University}
   \centerline{\it 3600 University Street, Montr\'eal}
   \centerline{Qu\'ebec, CANADA, H3A 2T8}
}


\def\anp#1#2#3{{\it Ann.~Phys. (NY)} {\bf #1} (19#2) #3}
\def\arnps#1#2#3{{\it Ann.~Rev.~Nucl.~Part.~Sci.} {\bf #1}, (19#2) #3}
\def\cmp#1#2#3{{\it Comm.~Math.~Phys.} {\bf #1} (19#2) #3}
\def\ijmp#1#2#3{{\it Int.~J.~Mod.~Phys.} {\bf A#1} (19#2) #3}
\def\jetp#1#2#3{{\it JETP Lett.} {\bf #1} (19#2) #3}
\def\jetpl#1#2#3#4#5#6{{\it Pis'ma Zh.~Eksp.~Teor.~Fiz.} {\bf #1} (19#2) #3
[{\it JETP Lett.} {\bf #4} (19#5) #6]}
\def\jpb#1#2#3{{\it J.~Phys.} {\bf B#1} (19#2) #3}
\def\mpla#1#2#3{{\it Mod.~Phys.~Lett.} {\bf A#1}, (19#2) #3}
\def\nci#1#2#3{{\it Nuovo Cimento} {\bf #1} (19#2) #3}
\def\npb#1#2#3{{\it Nucl.~Phys.} {\bf B#1} (19#2) #3}
\def\plb#1#2#3{{\it Phys.~Lett.} {\bf #1B} (19#2) #3}
\def\pla#1#2#3{{\it Phys.~Lett.} {\bf #1A} (19#2) #3}
\def\prc#1#2#3{{\it Phys.~Rev.} {\bf C#1} (19#2) #3}
\def\prd#1#2#3{{\it Phys.~Rev.} {\bf D#1} (19#2) #3}
\def\pr#1#2#3{{\it Phys.~Rev.} {\bf #1} (19#2) #3}
\def\prep#1#2#3{{\it Phys.~Rep.} {\bf C#1} (19#2) #3}
\def\prl#1#2#3{{\it Phys.~Rev.~Lett.} {\bf #1} (19#2) #3}
\def\rmp#1#2#3{{\it Rev.~Mod.~Phys.} {\bf #1} (19#2) #3}
\def\sjnp#1#2#3#4#5#6{{\it Yad.~Fiz.} {\bf #1} (19#2) #3
[{\it Sov.~J.~Nucl.~Phys.} {\bf #4} (19#5) #6]}
\def\zpc#1#2#3{{\it Zeit.~Phys.} {\bf C#1} (19#2) #3}


\global\nulldelimiterspace = 0pt


\def\goto{\mathop{\rightarrow}}
\def\gotoo{\mathop{\longrightarrow}}
\def\mapstoo{\mathop{\longmapsto}}

\def\df{\mathrel{:=}}
\def\fd{\mathrel{=:}}


\def\frac#1#2{{{#1} \over {#2}}\,}  
\def\hf{{1\over 2}}
\def\nth#1{{1\over #1}}
\def\sfrac#1#2{{\scriptstyle {#1} \over {#2}}}  
\def\stack#1#2{\buildrel{#1}\over{#2}}
\def\dd#1#2{{{d #1} \over {d #2}}}  
\def\ppartial#1#2{{{\partial #1} \over {\partial #2}}}  
\def\grad{\nabla}
\def\Square{{\vbox {\hrule height 0.6pt\hbox{\vrule width 0.6pt\hskip 3pt
        \vbox{\vskip 6pt}\hskip 3pt \vrule width 0.6pt}\hrule height 0.6pt}}}
\def\Dsl{\hbox{/\kern-.6700em\it D}} 
\def\dsl{\hbox{/\kern-.5300em$\partial$}}
\def\pxpsl{\hbox{/\kern-.5600em$p$}}
\def\ssl{\hbox{/\kern-.5300em$s$}}
\def\epssl{\hbox{/\kern-.5100em$\epsilon$}}
\def\delsl{\hbox{/\kern-.6300em$\nabla$}}
\def\lxpsl{\hbox{/\kern-.4300em$l$}}
\def\elxpsl{\hbox{/\kern-.4500em$\ell$}}
\def\kxpsl{\hbox{/\kern-.5100em$k$}}
\def\qxpsl{\hbox{/\kern-.5000em$q$}}
\def\transp#1{#1^{\sss T}} 
\def\sla#1{\raise.15ex\hbox{$/$}\kern-.57em #1}
\def\Pl{\gamma_{\sss L}}
\def\Pr{\gamma_{\sss R}}
\def\pwr#1{\cdot 10^{#1}}



\def\supsub#1#2{\mathstrut^{#1}_{#2}}
\def\sub#1{\mathstrut_{#1}}
\def\sup#1{\mathstrut^{#1}}
\def\rsub#1{\mathstrut_{\rm #1}}
\def\rsup#1{\mathstrut^{\rm #1}}

\def\twi{\widetilde}
\def\mybar{\bar}

\def\roughly#1{\mathrel{\raise.3ex\hbox{$#1$\kern-.75em\lower1ex\hbox{$\sim$}}}}
\def\lsim{\roughly<}
\def\gsim{\roughly>}

\def\bv#1{{\bf #1}}
\def\scr#1{{\cal #1}}
\def\op#1{{\widehat #1}}
\def\tw#1{\tilde{#1}}
\def\ol#1{\overline{#1}}



\def\bfa{{\bf a}}
\def\bfb{{\bf b}}
\def\bfc{{\bf c}}
\def\bfd{{\bf d}}
\def\bfe{{\bf e}}
\def\bff{{\bf f}}
\def\bfg{{\bf g}}
\def\bfh{{\bf h}}
\def\bfi{{\bf i}}
\def\bfj{{\bf j}}
\def\bfk{{\bf k}}
\def\bfl{{\bf l}}
\def\bfm{{\bf m}}
\def\bfn{{\bf n}}
\def\bfo{{\bf o}}
\def\bfp{{\bf p}}
\def\bfq{{\bf q}}
\def\bfr{{\bf r}}
\def\bfs{{\bf s}}
\def\bft{{\bf t}}
\def\bfu{{\bf u}}
\def\bfv{{\bf v}}
\def\bfw{{\bf w}}
\def\bfx{{\bf x}}
\def\bfy{{\bf y}}
\def\bfz{{\bf z}}


\def\Bfa{{\bf A}}
\def\Bfb{{\bf B}}
\def\Bfc{{\bf C}}
\def\Bfd{{\bf D}}
\def\Bfe{{\bf E}}
\def\Bff{{\bf F}}
\def\Bfg{{\bf G}}
\def\Bfh{{\bf H}}
\def\Bfi{{\bf I}}
\def\Bfj{{\bf J}}
\def\Bfk{{\bf K}}
\def\Bfl{{\bf L}}
\def\Bfm{{\bf M}}
\def\Bfn{{\bf N}}
\def\Bfo{{\bf O}}
\def\Bfp{{\bf P}}
\def\Bfq{{\bf Q}}
\def\Bfr{{\bf R}}
\def\Bfs{{\bf S}}
\def\Bft{{\bf T}}
\def\Bfu{{\bf U}}
\def\Bfv{{\bf V}}
\def\Bfw{{\bf W}}
\def\Bfx{{\bf X}}
\def\Bfy{{\bf Y}}
\def\Bfz{{\bf Z}}


\def\Sca{{\cal A}}
\def\Scb{{\cal B}}
\def\Scc{{\cal C}}
\def\Scd{{\cal D}}
\def\Sce{{\cal E}}
\def\Scf{{\cal F}}
\def\Scg{{\cal G}}
\def\Sch{{\cal H}}
\def\Sci{{\cal I}}
\def\Scj{{\cal J}}
\def\Sck{{\cal K}}
\def\Scl{{\cal L}}
\def\Scm{{\cal M}}
\def\Scn{{\cal N}}
\def\Sco{{\cal O}}
\def\Scp{{\cal P}}
\def\Scq{{\cal Q}}
\def\Scr{{\cal R}}
\def\Scs{{\cal S}}
\def\Sct{{\cal T}}
\def\Scu{{\cal U}}
\def\Scv{{\cal V}}
\def\Scw{{\cal W}}
\def\Scx{{\cal X}}
\def\Scy{{\cal Y}}
\def\Scz{{\cal Z}}


\def\Prob{\mathop{\rm Prob}}
\def\tr{\mathop{\rm tr}}
\def\Tr{\mathop{\rm Tr}}
\def\det{\mathop{\rm det}}
\def\Det{\mathop{\rm Det}}
\def\Log{\mathop{\rm Log}}
\def\Re{\rm Re\;}
\def\Im{{\rm Im\;}}
\def\diag#1{{\rm diag}\left( #1 \right)}


\def\bra#1{\langle #1 |}
\def\ket#1{| #1 \rangle}
\def\braket#1#2{\langle #1 | #2 \rangle}
\def\vev#1{\langle 0 | #1 | 0 \rangle}
\def\avg#1{\langle #1 \rangle}

\def\Bra#1{\left\langle #1 \right|}
\def\Ket#1{\left| #1 \right\rangle}
\def\Avg#1{\left\langle #1 \right\rangle}

\def\ddx#1#2{d^{#1}#2\,}
\def\ddp#1#2{\frac{d^{#1}#2}{(2\pi)^{#1}}\,}


\def\vacbra{{\bra 0}}
\def\vac{{\ket 0}}

\def\rhs{right-hand side}
\def\lhs{left-hand side}

\def\hc{{\rm h.c.}}
\def\cc{{\rm c.c.}}

\def\sm{standard model}
\def\smh{standard-model}
\def\km{Kobayashi Maskawa}
\def\kmh{Kobayashi-Maskawa}
\def\edm{e.d.m.}

\def\eV{{\rm \ eV}}
\def\keV{{\rm \ keV}}
\def\MeV{{\rm \ MeV}}
\def\GeV{{\rm \ GeV}}
\def\TeV{{\rm \ TeV}}

\def\cm{{\rm \ cm}}
\def\sec{{\rm \ sec}}

\def\ecm{{\it e}{\hbox{\rm -cm}}}

\voffset0.5truein


\def\ss{\scriptstyle}
\def\gem{$U_{\rm em}(1)$}
\def\em{{\rm em}}
\def\cc{{\rm cc}}
\def\nc{{\rm nc}}
\def\mw{M_{\sss W}}
\def\mz{M_{\sss Z}}
\def\gf{G_{\sss F}}
\def\gsm{g^{\sss SM}}
\def\hsm{h^{\sss SM}}
\def\rht{{\sss R}}
\def\lft{{\sss L}}
\def\gwk{$SU_\lft(2) \times U_Y(1)$}
\def\lsm{\Scl_{\rm SM}}
\def\leff{\Scl_{\rm eff}}
\def\lnew{\Scl_{\rm new}}
\def\lhat{\hat{\Scl}}
\def\lnewht{\hat{\Scl}_{\rm new}}
\def\twe{\tw{e}}
\def\twg{\tw{g}}
\def\twh{\tw{h}}
\def\twm{\tw{m}}
\def\sw{s_w}
\def\cw{c_w}
\def\Fhat{\hat{F}}
\def\Ahat{\hat{A}}
\def\fhat{\hat{f}}
\def\What{\hat{W}}
\def\Zhat{\hat{Z}}
\def\Ghat{\hat{G}}
\def\dpi{\delta \Pi}
\def\dgamma{\delta \Gamma}
\def\dlambda{\delta \Lambda}
\def\ssa{\gamma}
\def\ssv{{\sss V}}
\def\ssa{{\sss A}}
\def\ssw{{\sss W}}
\def\ssy{{\sss Y}}
\def\ssz{{\sss Z}}
\def\za{{\sss Z}\gamma}
\def\mw{M_{\sss W}}
\def\mz{M_{\sss Z}}
\def\smz{m_{\sss Z}}
\def\gf{G_{\sss F}}
\def\rht{{\sss R}}
\def\lft{{\sss L}}
\def\sm{{\sss SM}}
\def\tgv{{\sss TGV}}
\def\ww{{\sss WW}}
\def\aa{{\gamma \gamma}}
\def\zz{{\sss ZZ}}
\def\za{{\sss Z}\gamma}
\def\zg{{{\sss Z}\gamma}}
\def\lsm{\Scl_{\rm SM}}
\def\twe{\tw{e}}
\def\twmz{\tw{m}_\ssz}
\def\twmw{\tw{m}_\ssw}
\def\tws{\tw{s}_w}
\def\twc{\tw{c}_w}
\def\sw{s_w}
\def\cw{c_w}
\def\gl{g_\lft}
\def\gr{g_\rht}
\def\gwk{$SU_\lft(2) \times U_{\sss Y}(1)$}
\def\gz{g_\ssz}
\def\dgz{\Delta g_{1\ssz}}
\def\dgv{\Delta g_{1\ssv}}
\def\gv{g_\ssv}
\def\kv{\kappa_\ssv}
\def\dkv{\Delta \kv}
\def\dkz{\Delta \kappa_\ssz}
\def\dkg{\Delta \kappa_\gamma}
\def\lv{\lambda_\ssv}
\def\lz{\lambda_\ssz}
\def\lg{\lambda_\gamma}
\def\msbar{$\ol{\hbox{MS}}$}


\rightline{McGill-93/24}
\rightline{UdeM-LPN-TH-93-166}
\rightline{OCIP/C-93-9}
\rightline{July 1993}
\vskip .1in

\title
\centerline{A Global Fit to}
\centerline{Extended Oblique Parameters}
\endtitle

\authors
\centerline{C.P.~Burgess,${}^a$ Stephen Godfrey,${}^b$ Heinz K\"onig,
${}^{b}$\footnote{*}{\eightrm Address after September 1993: D\'epartement
de Physique,  l'Universit\'e du Qu\'ebec \'a Montr\'eal, C.P. 8888, Succ.
A, Montr\'eal, Qu\'ebec, CANADA, H3C 3P8.} David London${}^c$ and Ivan
Maksymyk${}^c$}
\vskip .15in
\centerline{\it ${}^a$ Physics Department, McGill University}
\centerline{\it 3600 University St., Montr\'eal, Qu\'ebec, CANADA, H3A
2T8.}
\vskip .1in
\centerline{\it ${}^b$Ottawa-Carleton Institute for Physics}
\centerline{\it Physics Department, Carleton University}
\centerline{\it Ottawa, Ontario, CANADA, K1S 5B6.}
\vskip .1in
\centerline{\it ${}^c$ Laboratoire de Physique Nucl\'eaire, l'Universit\'e
de Montr\'eal}
\centerline{\it C.P. 6128, Montr\'eal, Qu\'ebec, CANADA, H3C 3J7.}
\endauthors

\abstract
The $STU$ formalism of Peskin and Takeuchi is an elegant method for
encoding the measurable effects of new physics which couples to light
fermions dominantly through its effects on electroweak boson propagation.
However, this formalism cannot handle the case where the scale of new
physics is not much larger than the weak scale. In this case three new
parameters ($V, W$ and $X$) are required. We perform a global fit to
precision electroweak data for these six parameters. Our results differ
from what is found for just $STU$. In particular we find that the
preference for $S < 0$ is {\it not} maintained.
\endabstract


\vfill\eject
\section{Introduction}

As we impatiently await our first glimpse of physics beyond the standard
model, an important task is to develop methods for parametrizing measurable
effects of new physics. This activity constitutes the vital link between
experiment and the actual calculation of the effects of specific underlying
models. One such parametrization is the $STU$ treatment of Peskin and
Takeuchi
\ref\peskin{M.E. Peskin and T. Takeuchi, \prl{65}{90}{964};
\prd{46}{92}{381}; W.J. Marciano and J.L. Rosner, \prl{65}{90}{2963};
D.C. Kennedy and P. Langacker, \prl{65}{90}{2967}.}
\peskin, the end product of which is a set of expressions for electroweak
observables, consisting of a standard model prediction corrected by some
linear combination of the parameters $S$, $T$ and $U$. The $STU$ formalism
is applicable to the case of new physics which contributes to light-fermion
scattering dominantly through changes to the propagation of the electroweak
gauge bosons, and so is independent of the light-fermion generation. These
contributions are sometimes called `oblique' corrections. A wide variety of
models can be parametrized in this type of analysis, such as technicolor
models, multi-Higgs models, models with extra generations, and the like.

The $STU$ parametrization suffices when the scale of new physics $M$ is
large enough to justify approximating the new-physics contributions to
gauge-boson self energies at linear order in $q^2/M^2$. As is shown in
\ref\stuvwx{I. Maksymyk, C.P. Burgess and D. London, preprint McGill-93/13,
UdeM-LPN-TH-93-151, hepph-9306267 (unpublished).}
\stuvwx, however, the formalism breaks down when this approximation of
linearity fails, even if the dominant corrections are still of the oblique
form. Ref.~\stuvwx\ extends the $STU$ formalism to the case of general
oblique corrections, and shows that only three new parameters, $V$, $W$ and
$X$, are required to parametrize present data. The necessity for only six
parameters in all comes as something of a surprise, but is a consequence of
the present limitation of precision electroweak measurements to momentum
transfers $q^2 \approx 0$ and $q^2 = \mz^2$ or $\mw^2$.

The main situation for which the complete $STUVWX$ formalism is pertinent
is where the scale, $M$, of new physics is not large in comparison with the
weak scale. Comparatively light scalars and fermions arise in a great many
models, and can be potentially quite numerous in some of them, such as in
SUSY models for example.
Since these parameters are defined with an explicit factor of the
electromagnetic fine-structure constant, $\alpha$, and since the gauge
bosons couple  universally with strength of order $g = e/\sw$, the
contribution of any one of these light particles to the parameters $S$
through $X$ is expected to be of order $1/4\pi \sw^2 \simeq 0.3$. The
contribution of one or two electroweak multiplets of such states can be
expected to therefore contribute an amount that is comparable to 1.

The complete parametrization in terms of $S - X$ can also be required even
if the underlying scale happens to be large: $M \sim 1$ TeV. In this case
the more complete parametrization is necessary if the observables are to be
studied with a precision which is sensitive to $O(q^4/M^4)$
corrections\foot\nextterms{The subdominant terms in the $\ss q^2$
expansion have been
\ref\grinwise{B. Grinstein and M.B. Wise, \plb{265}{91}{326}.}
examined in Ref.~\grinwise.}. One instance where this accuracy is required
arises in the study of the loop-induced low-energy bounds on anomalous
electroweak boson self couplings
\ref\tgvanalysis{C.P. Burgess, S. Godfrey, H. K\"onig, D. London and I.
Maksymyk, preprint McGill-93/14, OCIP/C-93-7, UdeM-LPN-TH-93-154,
(unpublished).}
\tgvanalysis.

In this paper we perform a global fit to the current range of precision
electroweak measurements using this extended set of parameters. We do so
with two motivations in mind. First, we wish to determine the size of the
constraints that are implied by a joint fit for the parameters $S$ through
$X$. This permits the extraction of quantitative bounds on specific types
of new physics that do not satisfy the assumptions of the $STU$
parametrization, and indicates what kinds of models can be usefully
constrained in this way. Given their conventional normalization, we find
the parameters to be bounded to be $O(1)$.

\ref\negatives{For recent attempts to produce negative values for $S$ and
$T$ in underlying theories, see H. Georgi, \npb{363}{301}{1991};
E. Gates and J. Terning, \prl{67}{1840}{1991};
E. Ma and P. Roy, \prl{68}{92}{2879};
M. Luty and R. Sundrum, preprint LBL-32893-REV (unpublished);
L. Lavoura, L.-F. Li preprint DOE-ER-40682-27 (unpublished).}
Our second motivation is to see how the inclusion of $V$, $W$ and $X$
alters the previously-obtained bounds that have been obtained for $S$, $T$
and $U$ \peskin. In this case global fits tended to favour central values
for the parameter $S$ that were negative, with $S=1$ being excluded to the
$2\sigma$ level. This conclusion was particularly interesting considering
that many models of the underlying physics at scale $M$, such as
technicolour models, predict positive values for $S$ and $T$ \negatives.
Our more general fit finds that the preference for negative $S$ no longer
holds. In a joint fit for all six parameters we find that the $2\sigma$
allowed range for $S$ becomes $- 4.3 < S < 2.5$.

\section{Expressions for Observables in Terms of $S$ through $X$}

In so far as it is sufficient to encode new physics effects in gauge-boson
self energies only, one can express electroweak observables as the usual SM
prediction plus some linear combination involving new physics self energies
$\delta\Pi (q^2)_{ab}$, where $a, b = W,Z$, and $\gamma$. In this case, to
the extent that precision observables only probe $q^2 \approx 0$ and $q^2 =
\mz^2$ and $\mw^2$, it turns out that one can express all corrections to
electroweak observables in terms of six linearly independent combinations
of the various $\delta\Pi$'s. The contributions to linear order in $q^2$
may be parametrized by the three parameters $S$, $T$ and $U$, defined by
\label\stu
\eqa
{\alpha S \over 4 \sw^2 \cw^2 } &= \left[\, {\dpi_\zz(\mz^2) - \dpi_\zz(0)
\over \mz^2} \,\right] - {(\cw^2 - \sw^2) \over \sw\cw} \; \dpi_\zg'(0) -
\dpi'_\aa(0), \eol
\alpha T &= {\dpi_\ww(0) \over \mw^2} - {\dpi_\zz(0) \over \mz^2}, \eol
{\alpha U \over 4 \sw^2 } &= \left[\, {\dpi_\ww(\mw^2) - \dpi_\ww(0) \over
\mw^2} \, \right] - \cw^2 \left[ \, { \dpi_\zz(\mz^2) - \dpi_\zz(0) \over
\mz^2} \, \right] \eolnn
&\qquad  -  \sw^2 \dpi'_\aa(0) - 2 \sw \cw \dpi_\zg'(0).\eeol
\eeq
where the prime ($^\prime$) denotes differentiation with respect to $q^2$.
The remaining contributions may be subsumed into the new parameters $V$,
$W$ and $X$, defined as
\label\vwx
\eqa
\alpha V &= \dpi_\zz'(\mz^2) - \left[ \, {\dpi_\zz(\mz^2) - \dpi_\zz(0)
\over \mz^2}  \, \right], \eol
\alpha W &= \dpi_\ww'(\mw^2) - \left[ \, {\dpi_\ww(\mw^2) - \dpi_\ww(0)
\over \mw^2}  \, \right],  \eol
\alpha X &= - \sw \cw \left[ \, { \dpi_\zg(\mz^2) \over \mz^2} -
\dpi'_\zg(0) \, \right]. \eeol
\eeq
Manifestly, these expressions would vanish if $\dpi_{ab}(q^2)$ were simply
a linear function of $q^2$, in which the parametrization of new physics
effects could be achieved adequately with the set $STU$.

It is shown in Refs.~\peskin\ and \stuvwx\ how to calculate the dependence
of electroweak observables on the variables $S$ through $X$. In this
analysis, as is commonly done, we take as numerical inputs the following
three observables: $\alpha$ as measured in low-energy scattering
experiments, $\gf$ as measured in muon decay, and $\mz$. These observables
are chosen because they are the most precisely measured. With this choice,
the parameters $U$ appears only in the observables $\mw$ and $\Gamma_\ssw$,
and $W$ only appears in $\Gamma_\ssw$.\foot\bycontrast{By contrast, a
different choice of inputs -- such as $\ss \mz$, $\ss \mw$ and $\ss \alpha$
for instance -- would lead to $\ss U$-dependence throughout all the neutral
current observables.}

In observables defined at $q^2 \approx 0$, only the usual parameters $S$
and $T$ contribute. For example, the effective value of the weak mixing
angle, $(\sw^2)_{eff}$, as measured in various low-energy asymmetries,
(such as atomic parity violation, the low-energy neutral current scattering
ratio $R= {\sigma (\nu_\mu e)/ \sigma (\bar{\nu}_\mu e})$, \etc) is given
by
\label\seff \eq
(\sw^2)_{\rm eff}(q^2\! = \! 0) \;  = \; (\sw^2)_\sm
+ {\alpha S \over 4  ( \cw^2 - \sw^2)}
- { \sw^2 \cw^2 \; \alpha T \over \cw^2 - \sw^2}~,
\eeq
and the relative strength of the low-energy neutral- and charged-current
interactions is given by
\label\gammaz
\eq
\rho  = \rho_\sm(e,\gf,\mz) \; ( 1 + \alpha T ).
\eeq

As for measurements at the $Z$ resonance, the effective weak mixing angle
is given by
\eq
(\sw^2)_{eff}(q^2 \!=\! \mz^2) \;  =  \;
(\sw^2)_{\sss SM}(q^2 \!=\! \mz^2) + {\alpha \over 4 (\cw^2 - \sw^2)}S
 - {\cw^2\sw^2\alpha \over (\cw^2 -\sw^2)}T  + \alpha X ~, \eeq
and an example of a correction to $Z$-decay is
\eq
\Gamma (Z \rightarrow \bar{\nu}\nu ) \;=\; \Gamma_{\sss SM}(Z \rightarrow
 \bar{\nu}\nu )  \; ( 1 + \alpha T + \alpha V ) .\eeq
We thus see that $V$ describes a contribution to the overall normalization
of the strength of the neutral-current interaction, while $X$ acts to shift
the effective value of $(\sw^2)_{\rm eff}$ measured at the $Z$ pole.

The $W$ boson mass and width are
\label\wmass
\eqa
\mw^2 &= (\mw^2)_\sm(e,\sw,\smz) \left[ 1 - {\alpha S \over 2(\cw^2 -
\sw^2)} + {\cw^2 \alpha T \over (\cw^2 - \sw^2)} + {\alpha U \over 4 \sw^2}
\right], \eol
\Gamma (W \rightarrow all) &=  \Gamma_{\sss SM}(W \rightarrow all ) \;
\left[ 1 - {\alpha S\over 2 (\cw^2 - \sw^2)} - {\sw^2\alpha \over (\cw^2
-\sw^2)}T + {\alpha \over 4 \sw^2}U + \alpha W \right] . \eeol
\eeq
As advertised, the parameter $W$ turns out to appear only in the expression
for $\Gamma_\ssw$.

A comprehensive list of expressions for the electroweak observables that we
include in our analysis is given in Table I.  These expressions consist of
a radiatively corrected standard model prediction plus a linear combination
of the six parameters  $S$, $T$, $U$, $V$, $W$ and $X$. $\Gamma_\ssz$ and
$\Gamma_{b\bar{b}}$ are the total width and partial width into $b\bar{b}$;
$A_{\sss FB}(f)$ is the forward-backward asymmetry for $e^+e^- \to
f\bar{f}$; $A_{pol}(\tau)$, or $P_\tau$, is the polarization asymmetry
defined by $A_{pol}(\tau) = (\sigma_\rht - \sigma_\lft)/ (\sigma_\rht +
\sigma_\lft)$, where $\sigma_{\sss L,R}$ is the cross section for a
correspondingly polarized $\tau$ lepton; $A_e(P_\tau)$ is the joint
forward-backward/left-right asymmetry as normalized in
\ref\paul{P. Langacker, to appear in the {\sl Proceedings of 30 Years of
Neutral Currents}, Santa Monica, February 1993.}
Ref.~\paul;  and $A_{\sss L,R}$ is the polarization asymmetry which has
been measured by the SLD collaboration at SLC
\ref\sld{K. Abe \etal, \prl{70}{93}{2515}.}
\sld. The low-energy observables $g_\lft^2$ and $g_\rht^2$ are measured in
deep inelastic $\nu N$ scattering, $g^e_\ssv$ and $g^e_\ssa$ are measured
in $\nu e \to \nu e$ scattering, and $Q_\ssw(Cs)$ is the weak charge
measured in atomic parity violation in cesium.

\midinsert
$$\vbox{\tabskip=0pt \offinterlineskip
\halign to \hsize{\strut#& #\tabskip 1em plus 2em minus .5em&#\hfil
&#\tabskip=0pt\cr
\noalign{\hrule}\noalign{\smallskip}\noalign{\hrule}\noalign{\medskip}
&& \hfil \hbox{Expressions for Observables} &\cr
\noalign{\medskip}\noalign{\hrule}\noalign{\medskip}
&& $\Gamma_\ssz  =(\Gamma_\ssz)_{\sss SM} - 0.00961 S + 0.0263 T
 + 0.0194 V - 0.0207 X $  (GeV)  & \cr
&& $\Gamma_{b\ol{b}}  =(\Gamma_{b\ol{b}})_{\sss SM} - 0.00171 S + 0.00416 T
 + 0.00295 V - 0.00369 X$ (GeV)  & \cr
&& $\Gamma_{l^+ l^-}=(\Gamma_{l^+ l^-})_{SM} -0.000192 S
  + 0.000790 T + 0.000653 V - 0.000416 X $ (GeV)  &\cr
&& $\Gamma_{had}=(\Gamma_{had})_{SM} -0.00901 S
  + 0.0200 T + 0.0136 V - 0.0195 X $ (GeV) &\cr
&& $A_{\sss FB}(\mu)= (A_{\sss FB}(\mu))_{\sss SM} - 0.00677 S +
0.00479 T - 0.0146 X$  & \cr
&& $A_{pol}(\tau) = (A_{pol}(\tau))_{\sss SM} -0.0284 S + 0.0201 T - 0.0613
X $  & \cr
&& $A_e (P_\tau) =(A_e(P_\tau))_{\sss SM} -0.0284 S + 0.0201 T - 0.0613 X $
&  \cr
&& $A_{\sss FB}(b)=(A_{\sss FB}(b))_{\sss SM} -0.0188 S + 0.0131 T
-0.0406X$  & \cr
&& $A_{\sss FB}(c)=(A_{\sss FB}(c))_{\sss SM} -0.0147 S + 0.0104 T -0.03175
X$  &\cr
&& $A_{\sss LR} =(A_{\sss LR})_{\sss SM} -0.0284 S + 0.0201 T - 0.0613 X  $
& \cr
&& $M_\ssw^2=(M_\ssw^2)_{\sss SM}(1-0.00723 S +0.0111 T +0.00849 U)$ &\cr
&& $\Gamma_\ssw =(\Gamma_\ssw)_{\sss SM}(1-0.00723 S -0.00333 T + 0.00849 U
+ 0.00781W) $ & \cr
&& $g_{\sss L}^2 =(g_{\sss L}^2)_{\sss SM}-0.00269 S + 0.00663 T$ & \cr
&& $g_{\sss R}^2 =(g_{\sss R}^2)_{\sss SM} +0.000937 S - 0.000192  T$ & \cr
&& $g_{\sss V}^e(\nu e \to \nu e) =(g_{\sss V}^e)_{\sss SM} + 0.00723 S -
0.00541 T$  & \cr
&& $g_{\sss A}^e (\nu e \to \nu e) =(g_{\sss A}^e)_{\sss SM} - 0.00395 T$ &
\cr
&& $Q_\ssw(^{133}_{55}Cs) = Q_\ssw(Cs)_{\sss SM} -0.795 S -0.0116 T$ & \cr
\noalign{\medskip}\noalign{\hrule}\noalign{\smallskip}\noalign{\hrule}
}}$$
\centerline{{\bf TABLE I}}
\medskip
\noindent {\eightrm Summary of the dependence of electroweak observables on
$\ss S,T,U,V,W$ and $\ss X$. In preparing this table we used the numerical
values $\ss \alpha(\mz^2)=1/128$ and $\ss \sw^2=0.23$.}
\endinsert

There are several features in Table I worth pointing out. First, as has
already been mentioned, due to the choice of numerical inputs ($\alpha$,
$\gf$, $\mz$), only the two parameters $S$ and $T$ contribute to the
observables for which $q^2\sim 0$; the parameter $U$ appears only in $\mw$
and $\Gamma_\ssw$.
The limit on $U$ comes principally from the $\mw$ measurement, since
$\Gamma_\ssw$ is at present comparatively poorly measured. For the same
reason, the parameter $W$ is weakly bounded, since it contributes only to
$\Gamma_\ssw$. In addition to $S$ and $T$, observables on the $Z^0$
resonance are also sensitive to $V$ and $X$, which are expressly defined at
$q^2=\mz^2$. Observables that are not explicitly given in Table I can be
obtained using the given expressions. In particular the parameter $R$ is
defined as $R= \Gamma_{had}/\Gamma_{l\bar{l}}$, and $\sigma^h_p =
12\pi\Gamma_{e\bar{e}}\Gamma_{had}/\mz^2\Gamma_\ssz^2$ is the hadronic
cross section at the $Z$-pole.

\section{Numerical Fit of $STUVWX$}

We now determine the phenomenological constraints on $STUVWX$ by performing
a global fit to the precision data. The experimental values and standard
model predictions of the observables used in our fit are given in Table II.
The standard model predictions are taken from
\ref\smpredictions{The standard model predictions come from: P. Langacker,
Proceedings of the 1992 Theoretical Advanced Study Institute, Boulder CO,
June 1992, which includes references to the original literature. We thank
P. Turcotte for supplying us with the standard model values for $g^2_{\sss
L}$ and $g^2_{\sss R}$.}
Ref.~\smpredictions\ and have been calculated using the values $m_t=150$
GeV and $M_H=300$ GeV. The LEP observables in Table II were chosen because
they are closest to what is actually measured, and are relatively weakly
correlated. In our analysis we include the combined LEP values for the
correlations
\ref\correlations{The LEP Collaborations: ALEPH, DELPHI, L3, and OPAL,
\plb{276}{92}{247}.}
\correlations.
\ref\lep{C. DeClercqan, Proceedings of the Recontre de Moriond, Les
Arcs France, March 1993; V. Innocente, {\it ibid}. }

\pageinsert
$$\vbox{\tabskip=0pt \offinterlineskip
\halign to \hsize{\strut#& #\tabskip 1em plus 2em minus .5em&\hfil#\hfil
&\hfil#\hfil &\hfil#\hfil &#\tabskip=0pt\cr
\noalign{\hrule}\noalign{\smallskip}\noalign{\hrule}\noalign{\medskip}
&& \hfil \hbox{Quantity} & \hfil \hbox{Experimental Value}&
\hfil \hbox{Standard Model Prediction} &\cr
\noalign{\medskip}\noalign{\hrule}\noalign{\medskip}
&& $\mz$ (GeV)  & $91.187 \pm 0.007 $ \lep & input & \cr
&& $\Gamma_\ssz$ (GeV)  & $ 2.488 \pm 0.007 $ \lep & $2.490 [\pm0.006]$ &
\cr
&& $R=\Gamma_{had}/\Gamma_{l{\bar l}}$  & $20.830 \pm 0.056$ \lep
	& $20.78 [\pm 0.07]$ & \cr
&& $\sigma^h_p $ (nb)  & $41.45 \pm 0.17$ \lep & $41.42 [\pm 0.06]$ & \cr
&& $\Gamma_{b\bar b}$ (MeV)  & $383 \pm 6$ \lep & $ 375.9 [\pm 1.3]$ & \cr
&& $A_{FB}(\mu)$ & $0.0165 \pm 0.0021$ \lep & $0.0141$ & \cr
&& $A_{pol}(\tau)$  & $0.142 \pm 0.017$ \lep & $ 0.137$ & \cr
&& $A_e (P_\tau)$  & $ 0.130 \pm 0.025$ \lep & $ 0.137$ & \cr
&& $A_{FB}(b)$  & $0.0984 \pm 0.0086 $ \lep & $0.096$ & \cr
&& $A_{FB}(c)$  & $ 0.090 \pm 0.019$ \lep & $ 0.068$ &\cr
&& $A_{LR} $  & $ 0.100 \pm 0.044$ \sld & $ 0.137$ & \cr
\noalign{\medskip}\noalign{\hrule}\noalign{\medskip}
&& $\mw$ (GeV)  & $79.91 \pm 0.39$
\ref\cdf{R. Abe {\sl et al.}, \prl{65}{90}{2243}.}\cdf & $80.18$ &\cr
&& $\mw/\mz$ 	& $0.8798 \pm 0.0028 $
\ref\uatwo{J. Alitti {\sl et al.},
	\plb{276}{92}{354}.}\uatwo & 0.8793 & \cr
&& $\Gamma_\ssw$  (GeV) & $2.12 \pm 0.11$
\ref\pdb{Particle Data Group, \prd{45}{92}{II}.}\pdb & 2.082 & \cr
\noalign{\medskip}\noalign{\hrule}\noalign{\medskip}
&& $g_\lft^2 $  & $0.3003 \pm 0.0039$
\paul & $0.3021$ & \cr
&& $g_\rht^2 $  & $0.0323\pm 0.0033 $ \paul & $0.0302$ & \cr
&& $g_\ssa^e $  & $ -0.508 \pm 0.015 $ \paul & $-0.506$ & \cr
&& $g_\ssv^e $  & $ -0.035\pm 0.017 $ \paul & $-0.037$ & \cr
&& $Q_\ssw(Cs)$ & $-71.04 \pm 1.58 \pm [0.88]$
\ref\cesium{M.C. Noecker {\sl et al.},
	\prl{61}{88}{310}.}\cesium & $-73.20$ & \cr
\noalign{\medskip}\noalign{\hrule}\noalign{\smallskip}\noalign{\hrule}
}}$$
\centerline{{\bf TABLE II}}
\medskip
\noindent {\eightrm Experimental values for electroweak observables
included in global fit. The $\ss Z$-pole measurements are the preliminary
1992 LEP results taken from Ref.~\lep. The couplings extracted from
neutrino scattering data are the current world averages taken from
Ref.~\paul. The values for standard model predictions are taken from
Ref.~\smpredictions\ and have been calculated using $\ss m_t=150$ GeV and
$\ss M_{\sss H}=300$ GeV. We have not shown the errors in the standard
model predictions associated with theoretical uncertainties in radiative
corrections or with the uncertainty regarding the measurement of $\ss \mz$,
since these errors are in general overwhelmed by experimental errors. The
exception is the error due to uncertainty in $\ss \alpha_s$, shown in
square brackets. We include this error in quadrature in our fits. The error
in square brackets for $\ss Q_\ssw(Cs)$ reflects the theoretical
uncertainty regarding atomic wavefunctions
\ref\atomictheory{S.A. Blundell, W.R. Johnson, and J. Sapirstein,
\prl{65}{90}{1411}; V.A. Dzuba {\sl et al.}, \pla{141}{89}{147}.}
\atomictheory\ and is also included in quadrature with the experimental
error.}
\endinsert

In Table III are displayed the results of the fit. In the second column are
shown the results of individual fits, obtained by setting all but one
parameter to zero. The third column is a fit of $STU$, with $VWX$ set to
zero. Finally, in column four, we give the results for the fit in which all
six parameters were allowed to vary simultaneously.

\midinsert
$$\vbox{\tabskip=0pt \offinterlineskip
\halign to \hsize{\strut#& #\tabskip 1em plus 2em minus .5em&\hfil#\hfil
&\hfil#\hfil &\hfil#\hfil &\hfil#\hfil &#\tabskip=0pt\cr
\noalign{\hrule}\noalign{\smallskip}\noalign{\hrule}\noalign{\medskip}
&& \hfil \hbox{Parameter} & \hfil \hbox{Individual Fit}&
\hfil \hbox{STU Fit} &  \hfil \hbox{STUVWX Fit}  &\cr
\noalign{\medskip}\noalign{\hrule}\noalign{\medskip}
&& $S$ & $-0.19 \pm 0.20 $ & $-0.48\pm 0.40$ & $-0.93 \pm 1.7$   & \cr
&& $T$ & $0.06  \pm 0.19 $ & $-0.32\pm 0.40$ & $-0.67 \pm 0.92$  & \cr
&& $U$ & $-0.12 \pm 0.62 $ & $-0.12\pm 0.69$ & $-0.6 \pm 1.1$  & \cr
&& $V$ & $-0.09 \pm 0.45 $ & --- & $0.47\pm 1.0$  & \cr
&& $W$ & $2.3   \pm 6.8 $  & --- & $1.2\pm 7.0$    & \cr
&& $X$ & $-0.10 \pm 0.10 $ & --- & $0.10\pm 0.58$  & \cr
\noalign{\medskip}\noalign{\hrule}\noalign{\smallskip}\noalign{\hrule}
}}$$
\centerline{{\bf TABLE III}}
\medskip
\noindent {\eightrm Global fits of $\ss STUVWX$ to precision electroweak
data. The second column contains the results of inidividual fits, obtained
by setting all but one parameter to zero. The third column is a fit of $\ss
STU$ setting $\ss VWX$ equal to zero, and the final column allows all
parameters to vary simultaneously. We have shown the 1$\ss \sigma$ errors.}
\endinsert

The most important observation concerning these results is that all of the
parameters are consistent with zero. In other words there is no evidence
for physics outside the standard model. The second observation is that
including $VWX$ in our fits weakens the constraints on $STU$. This can be
seen graphically in Fig.~1 where we have plotted the 68\% and 90\% C.L.
contours for $S$ and $T$. We show the results for the case in which the
parameters $VWX$ have  been set to zero as well as that in which they have
been allowed to vary. Notice in particular that while the entire $1-\sigma$
allowed range for $S$ in the $STU$ fit satisfies $S<0$, this is not true
for the fit with all six parameters.

\section{Conclusions}

We have performed a global fit for the complete set of six oblique
correction parameters, $S$ through $X$. This fit extends the results of
previous fits for $S$, $T$ and $U$ to a much wider class of models for the
underlying physics, including in particular new light particles which need not
be much heavier that the weak scale. We find that these parameters are bounded
by the data to be
$\lsim 1$, corresponding to an $O(1\%)$ correction to the weak-boson vacuum
polarizations, $\delta \Pi(q^2)$. Such bounds are sensitive enough to
constrain many models for new physics near the weak scale, much as did the
original $STU$ analysis for technicolour models at the TeV scale.

We have also compared our joint fit of the six parameters $S$ through $X$
to a three-parameter fit involving only $S$, $T$, and $U$ (with
$V\!=\!W\!=\!X\!=\!0$). Not surprisingly, we find in the general case that
the allowed ranges for $S$, $T$, and $U$ are relaxed. In particular, the
preference found in earlier fits for negative values for $S$ -- which had
been uncomfortable for many underlying models -- are no longer present for
the six-parameter fit.

\bigskip
\centerline{\bf Acknowledgements}
\bigskip
S.G. and D.L. gratefully acknowledge helpful conversations and
communications with Paul Langacker, and thank Paul Turcotte for supplying
standard model values for $g^2_{\sss L}$ and $g^2_{\sss R}$. This research
was partially funded by funds from the N.S.E.R.C.\ of Canada and le Fonds
F.C.A.R.\ du Qu\'ebec.

\bigskip
\bigskip
\centerline{\bf Figure Captions}
\bigskip

\topic{Figure 1} Constaints on $S$ and $T$ from a global fit of precison
electroweak measurements. The solid line represents the 68\% C.L. setting
$VWX$ to zero, the dashed line represents the 90\% C.L. setting $VWX$ to
zero, the dotted line represents the 68\% C.L. allowing $VWX$ to vary, and
the dot-dashed line represents the 90\% C.L. allowing $VWX$ to vary.

\listrefs

\bye